\documentclass[twocolumn,superscriptaddress,showpacs,preprintnumbers,amsmath,amssymb,prl]{revtex4}
\usepackage{graphicx}
\usepackage{color}
\usepackage{dcolumn}
\usepackage{bm}

\begin{document}

\title{Synchronization dynamics in a designed open system}
\author{Nobuhiko Yokoshi}
\email{yokoshi@pe.osakafu-u.ac.jp}
\affiliation{Department of Physics and Electronics, Osaka Prefecture University, Sakai, Osaka 599-8531, Japan}
\author{Kazuki Odagiri}
\affiliation{Department of Physics and Electronics, Osaka Prefecture University, Sakai, Osaka 599-8531, Japan}
\author{Akira Ishikawa}
\affiliation{Department of Science for Advanced Materials, University of Yamanashi, Kofu 400-8511, Japan} 
\author{Hajime Ishihara}
\affiliation{Department of Physics and Electronics, Osaka Prefecture University, Sakai, Osaka 599-8531, Japan}
\affiliation{} 

\date{\today}

\begin{abstract}
We theoretically propose a unifying expression for synchronization dynamics between two-level constituents. Although synchronization phenomena require some substantial mediators, the distinct repercussions of their propagation delays remain obscure, especially in open systems. Our scheme directly incorporates the details of the constituents and mediators in an arbitrary environment. As one example, we demonstrate the synchronization dynamics of optical emitters on a dielectric microsphere. We reveal that the whispering gallery modes (WGMs) bridge the well-separated emitters and accelerate the synchronized fluorescence, known as superfluorescence. The emitters are found to overcome the significant and nonuniform retardation, and to build up their pronounced coherence by the WGMs, striking a balance between the roles of resonator and intermediary. Our work directly illustrates the dynamical aspects of many-body synchronizations and contributes to the exploration of research paradigms that consider designed open systems.
\end{abstract}
\maketitle
Since Huygens noticed that two pendulum clocks with a common support tend to exhibit synchronized oscillation~\cite{huygens}, spontaneous synchronization without external driving has been recognized as ubiquitous behavior appearing in many areas of science and engineering~\cite{coope, clocks,engineering,engineering2}. The concept was extended to quantum mechanical systems such as Josephson junction and laser arrays~\cite{coope}. One of the most exemplary targets is the synchronization of photoemissions, i.e., superfluorescence: a particular style of Dicke superradiance~\cite{Dicke}. Densely spaced dipoles exchange photons and radiate a pulsed light after building up a macroscopic polarization.  Such dynamics have been observed in atomic gases~\cite{gSF1,gSF2}, and impurities or carriers in solids~\cite{sSF1,sSF2,sSF3,sSF4}. The recent realization in artificial nanostructures~\cite{noell,miyajima} is a development that is promising to both fundamental many-body studies and applications. The conceivable next task is to design the synchronization by controlling dipole configuration and the surrounding environment. Despite numerous studies from various approaches~\cite{tSFr,tSF1,tSF2,PD1,PD2,de2,ishikawa,Zhu,Martin}, no practical scheme to approach this problem has yet been established.

As every synchronization is mediated by external agents, there must be some retardation effect. More importantly, the mediators themselves are scattered, absorbed, or amplified in open systems. Therefore, for a practical and versatile evaluation of the dynamics, it is essential to incorporate these effects. Here, we present a synchronization equation for two-level optical emitters that directly incorporates their locations and the mediators' nonuniform propagations. The proposed scheme makes it possible to treat the emitters in natural or artificial lattices, as well as in optical resonators~\cite{Vahala} or plasmonic near fields~\cite{Novotny}. To design the synchronizing system in detail gives us an actual advantage in building original functions of many-body systems.

Let us consider the dynamics of $N$ two-level emitters. The emitter at ${\bf r}={\bf r}_i$ has the dipole moment ${\bf d}_i=d~{\bf e}_{i}~(i=1, 2, \cdots, N),$ with ${\bf e}_{i}$ being the unit vector. The photon configures the electric field ${\bf E}({\bf r})=i\Sigma_{\alpha} \int d{\bf k} \sqrt{\frac{\hbar c |{\bf k}|}{16\pi^3\epsilon_0}} b_{{\bf k},\alpha} {\bf f}_{{\bf k},\alpha}({\bf r})+{\rm H.c.}$, where the operator $b_{{\bf k},\alpha}$ annihilates a photon of the wave function $ {\bf f}_{{\bf k},\alpha}({\bf r})$ having the wave vector ${\bf k}$ and polarization $\alpha=\pm1$. Here, $\hbar$ is the Dirac constant, $c$ the speed of light, and $\varepsilon_0$ the vacuum permittivity. The whole Hamiltonian is
\begin{eqnarray}
H&=&\sum_{\alpha} \int d{\bf k} \hbar c|{\bf k}| b_{\bf k, \alpha}^{\dagger}b_{\bf k, \alpha} \nonumber \\
&&+\sum_{i=1}^N \hbar \omega_i \sigma_i^+ \sigma_i^- 
-\sum_{i=1}^N {\bf d}_i \cdot {\bf E}({\bf r}_i) (\sigma_i^+ +\sigma_i^-),
\end{eqnarray}
where the operator $\sigma_i^{\pm}$  flips the two levels separated by the frequency $\omega_i$. Assuming that the light-matter interaction is near resonant and much smaller than the intrinsic energies, we safely neglect $b_{{\bf k},\alpha}^{\dagger} \sigma_i^+$ and $b_{{\bf k},\alpha} \sigma_i^-$.

To estimate the emitter correlation, we aim to incorporate the photon field by introducing the retarded photon Green function, instead of the wavefunction itself. Hence, we simultaneously solve the Heisenberg equations for the following three density matrix elements: photon density $\langle b_{\bf k, \alpha}^{\dagger}b_{\bf k, \alpha}  \rangle$, photon-assisted polarization $\langle b_{\bf k, \alpha}^{\dagger}\sigma_i^-  \rangle$, and emitter-emitter correlation $\langle \sigma_i^+ \sigma_j^-  \rangle$. Although a full quantum mechanical treatment causes an infinite hierarchy of the equations, we can truncate this series using dynamical decoupling and treat up to the second-order cumulant of $\sigma_i^{\pm}$~\cite{kira,koch}. Note that the expectation values of $\langle b_{\bf k, \alpha} \rangle$ and $\langle \sigma_i^{\pm} \rangle$ are disregarded as we do not consider a coherent pumping. Here we apply the adiabatic approximation to the simultaneous equations; the time derivative of $\langle b_{\bf k, \alpha}^{\dagger}\sigma_i^-  \rangle$ is taken to be zero, leading to the stationary solution $ \langle b_{{\bf k},\alpha}^{\dagger} \sigma_i^- \rangle \propto i\Sigma_{m=1}^N \langle \sigma_m^+ \sigma_i^- \rangle {\bf d}_i\cdot {\bf f}_{{\bf k},\alpha}({\bf r}_i)/(c|{\bf k}|-\omega_i)$. This approximation is justified in typical excitonic systems, as the formation of the polarization is much faster than the other time scales. The validity of this approximation was intensively discussed in the previous study~\cite{ishikawa}. Substituting this solution into the simultaneous equations, the equation for $\langle \sigma_i^+ \sigma_j^-  \rangle$ becomes independent of the photon operator $b_{{\bf k}, \alpha}$, and the photon field appears only through the dyadic Green function,
\begin{eqnarray*}
\bar{ G}({\bf r},{\bf r}',\omega)=\sum_{\alpha}\int d{\bf k} 
\frac{c |{\bf k}|  \bigl[ {\bf f}_{{\bf k},\alpha}({\bf r}) \otimes {\bf f}^*_{{\bf k},\alpha}({\bf r}') \bigr]}{16\pi^3 (c |{\bf k}|-\omega)},
\end{eqnarray*}
which characterizes the profile of the electric field ${\bf E}({\bf r})$ in the absence of the emitters, and includes both the longitudinal and transverse components~\cite{wubs}. 

As a result, the emitter-emitter correlation is found to obey the equation
\begin{eqnarray}
&&\frac{\partial}{\partial t} \langle  \sigma_i^+(t) \sigma_j^-(t) \rangle
=
i\left ( \omega _i-\omega _j \right )
 \langle  \sigma_i^+ \sigma_j^- \rangle 
\nonumber\\
&&
+\frac{i}{2}
\sum^{N}_{m= 1}
\left[
 \langle  \sigma_i^+ \sigma_m^- \rangle \frac{C_{mj}}{\tau_m} \langle \sigma_j^z  \rangle
-
\langle \sigma_i^z  \rangle \frac{C_{im}}{\tau_i} \langle \sigma_m^+ \sigma_j^- \rangle
\right],
\label{kuramoto}
\end{eqnarray}
where $\sigma_i^z=(\sigma_i^+\sigma_i^--\sigma_i^-\sigma_i^+)$ represents the excited state occupation. The emitter-emitter coupling is determined by the radiation decay time $\tau_i=3\pi c^3 \hbar \varepsilon_0 / (\omega_i^3 d^2)$~\cite{Loudon}. The dimensionless function $C_{ij}=(6\pi c^3/\omega_i^3){\bf e}_{i} \cdot \bar{G}({\bf r}_i,{\bf r}_j, \omega_i) \cdot {\bf e}_{j}$ describes the retarded propagator between the distant emitters. When all the dipole moments are aligned and closed packed in a small region, no retardation appears (${\rm Im} [ C_{ij} ] \rightarrow 1$)~\cite{cutoff}. Moreover, the equation becomes equivalent to the Kuramoto model for the closed system; i.e., $|\langle \sigma_i^{+} (t) \sigma_j^{-} (t)\rangle|=1$~\cite{coope}.

There exist various factors that prevent the synchronization from being met~\cite{coope}. Among them, the retardation effect can be a limiting factor since the surrounding environment itself confines or radiates the photon with time. Such a nonequilibrium retardation is quite sensitive to the locations of the emitters because the emitted photons do not always propagate uniformly. Here, we have incorporated the photon propagations directly into the time-evolution equation of the emitter-emitter correlations. Once the photon Green function is obtained analytically or numerically, we can examine how the emitters build up the cooperative correlation for an arbitrary environment, locations, initial occupancies, and intrinsic frequencies.  From that perspective, Eq.~(\ref{kuramoto}) provides a unifying framework to simulate synchronizations, and helps to design a synchronizing system. Furthermore, as far as the spin-boson model represents the whole system~\cite{sbm}, it may be applied without any change to synchronizations utilizing other mediators: phonons~\cite{saw1,saw2}, magnons~\cite{mag1,mag2}, and plasmons~\cite{pls}.

\begin{figure}[tb]
\includegraphics[width=80mm]{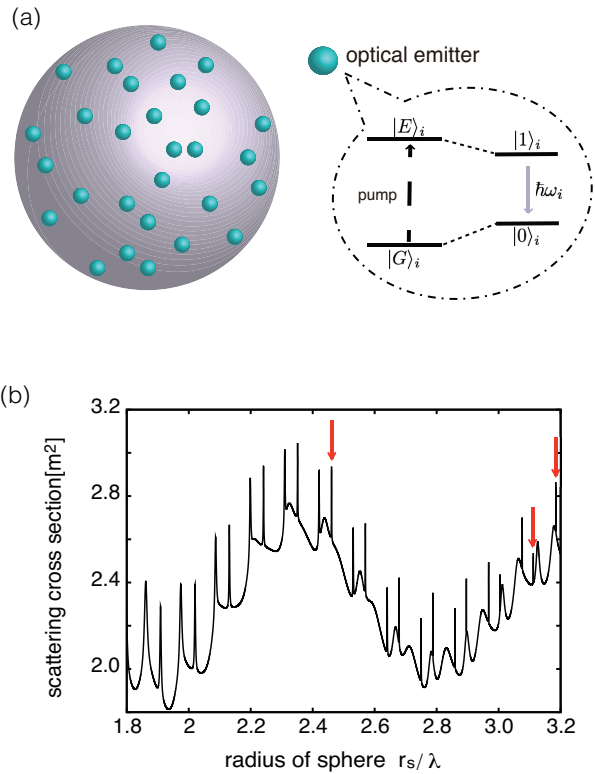}
\caption{(a) Schematic view of four-level optical emitters randomly positioned on a polystyrene sphere, and an energy diagram of them. We assume that all the dipoles are aligned in one direction. Utilizing the pump light resonating with none of the WGMs, we first excite the higher state (or continuum) $|E\rangle_i$, which relaxes into the state $|1\rangle_i$ by fast and radiationless decay. This prepares the initial state, whereby each two-level state \{$|0\rangle_i, |1\rangle_i$\} is incoherent and population inverted. (b) Scattering cross section of the polystyrene sphere plotted against its radius $r_s$. Each peak corresponds to the WGM that resonates with the $|0\rangle_i$-$|1\rangle_i$ excitations. }
\label{model} 
\end{figure}
In the following, we demonstrate the utility of our scheme in the specific system. We focus on uniaxial two-level emitters that are positioned randomly on a dielectric polystyrene microsphere [Fig.~\ref{model}(a)]. Such microspheres usually act as whispering gallery mode (WGM) resonators when photons propagate around the boundary by total internal reflection~\cite{WGM}. However, the smaller the sphere radius becomes, the faster the microsphere radiates WGMs through quantum tunneling effects at the curved boundary. Therefore, such modes can bridge the coherence of more than one emitter. To visualize the synchronization, we calculate the fluorescent dynamics $I(t) \equiv \sum_{\alpha}\int d{\bf k} \partial \langle b^{\dagger}_{{\bf k},\alpha }b_{{\bf k},\alpha} \rangle /\partial t$. The time derivative of the photon density obeys the Heisenberg equation: $\partial \langle b_{\bf k, \alpha}^{\dagger}b_{\bf k, \alpha}  \rangle/\partial t=2\Sigma_{i=1}^N {\rm Re}[ \sqrt{\frac{\hbar c |{\bf k}|}{16\pi^3\epsilon_0}} \langle b_{{\bf k},\alpha}^{\dagger}\sigma_i^- \rangle {\bf f}_{{\bf k},\alpha}({\bf r}_i)\cdot {\bf d}_i]$. Using the adiabatic approximation as in deriving Eq.~(\ref{kuramoto}), one can find the expression
\begin{eqnarray}
I(t)=\sum_{i,j=1}^N \frac{1}{\tau_i}  {\rm Im} \left[  C_{ij} \langle  \sigma_i^+(t) \sigma_j^- (t)\rangle \right].
\label{int}
\end{eqnarray}
It becomes equivalent to the well-known expression of the fluorescence when no retardation appears (${\rm Im}[C_{ij}] \rightarrow 1$)~\cite{tSFr}. The specific form of the Green function is obtained by matching the solutions of Maxwell's equation, $[ \nabla \times \nabla \times -\epsilon({\bf r})(\omega^2/c^2)]  \bar{G}({\bf r},{\bf r}',\omega) =(\omega^2/c^2) \delta({\bf r}-{\bf r}')\bar{\bf 1},$ at the boundary~\cite{tai}. In this study, we analytically connect the Green function inside the microsphere [the relative permittivity $\epsilon({\bf r})=2.5$] and that in a vacuum [$\epsilon({\bf r})=1$]. The Green function essentially includes the divergence at ${\bf r}={\bf r}'$. Thus, we have to introduce an ultraviolet cutoff~\cite{cutoff}. Here, we have employed the cutoff originating from the radius of the emitters, $\sim 5{\rm nm}$. 

We assume that all the emitters are initially in the excited state and are uncorrelated, i.e., $\langle \sigma_i^+ (0) \sigma_j^- (0) \rangle=\delta_{ij}$, and that the microsphere is in its ground state. To prepare such a state, the four-level molecules are considered to be prospective emitters (Fig.~\ref{model}(a)); the intermediate states \{$|0\rangle_i, |1\rangle_i$\} act as two-level emitters that are separated by a common energy $\hbar \omega=3.0 {\rm eV}$ (corresponding to the wavelength $\lambda=413 {\rm nm}$). For simplicity, we assume that their dipole moments are aligned and that their radiation delay times are the same $\tau=100 {\rm ps}$. The scattering cross section of the microsphere is calculated by Mie scattering theory~\cite{tai}. In Fig.~\ref{model}(b), we plot this against the sphere radius $r_s$ divided by $\lambda$. The resonant WGMs are identified by the sharp peaks, among which we choose the radii indicated by red arrows. When the sphere radius is $r_s/\lambda= 3.184$, the corresponding WGM is a transverse magnetic mode with the lifetime $\tau_{\rm WGM}= 40{\rm ps}$. Additionally, for $r_s/\lambda= 3.112$ and $2.461$, the corresponding WGMs are the transverse electric modes whose lifetimes are  $\tau_{\rm WGM}=20$ and $6{\rm ps}$, respectively. The polarization of the WGM does not clearly influence the fluorescent dynamics due to the randomness of the uniaxial emitter positions. On the other hand, its lifetime can change the fluorescent statistics~\cite{de1}.

\begin{figure}[tb]
\includegraphics[width=70mm]{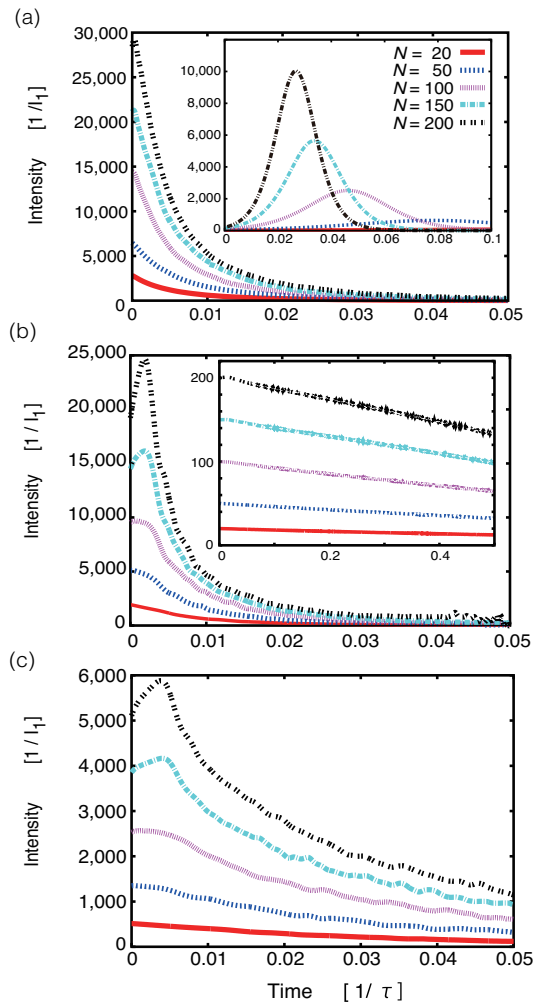}
\caption{Fluorescent dynamics from the optical emitters. Here, $I_1=1/\tau$ is the fluorescence intensity from an isolated emitter in a vacuum. (a) Spontaneous emission accelerated by the resonator effect of the WGM in the polystyrene sphere ($r_s/\lambda =3.184$). Inset: We plot the synchronized fluorescence (superfluorescence) where all the emitters are close packed in a vacuum. (b) Synchronized fluorescence when the polystyrene sphere radii is $r_s/\lambda=3.112$. Inset: We plot the intensity when the microsphere is removed. (c) The same plot for the case using the sphere with $r_s/\lambda=2.461$.  }
\label{wgm} 
\end{figure}
Figure~\ref{wgm}(a) shows the fluorescent dynamics from the emitters on the sphere $r_s/\lambda =3.184$, where the ratio of the radiation decay time of the emitters and that of the corresponding WGM is  $\eta=\tau_{\rm WGM}/\tau=0.4$. Although the observed intensities are enhanced, they exhibit the exponential decays, i.e., $I(t) \sim I(0)\exp[ -I(0) t /N]$. This has no resemblance to the Dicke superfluorescence~\cite{Dicke}; it is reproduced when all the emitters are closed packed in a vacuum [see inset of Fig.~\ref{wgm}(a)]. This means that the polystyrene sphere acts exactly as a resonator. The photons are confined in the sphere as the WGMs rather than mediating the emitters.

Utilizing just a little smaller sphere, the situation changes dramatically. Figures~\ref{wgm}(b) and \ref{wgm}(c) show the fluorescent dynamics for $r_s/\lambda =$3.112 ($\eta=0.2$) and 2.461 ($\eta=0.06$). One can see that the intensities increase at the onset of the fluorescence for $N \ge 100$, which clearly indicates the appearance of the synchronization. It should be noted that the mean separation between the emitters exceeds the radiation wavelength $\lambda$. The mediator photons distinguish the locations of the individual emitters sufficiently, and the retardation effect can be serious. However, since all the WGMs propagate while geometrically confined near the spherical surface, the phase disturbance due to the random retardations is alleviated as a whole. In addition, as the lifetime $\tau_{\rm WGM}$ decreases, the WGM penetrates more deeply into the vacuum, and thus the interemitter communication becomes activated. In fact, the synchronization disappears when only the sphere is removed [see inset of Fig.~\ref{wgm}(b)].  
\begin{figure}[tb]
\includegraphics[width=70mm]{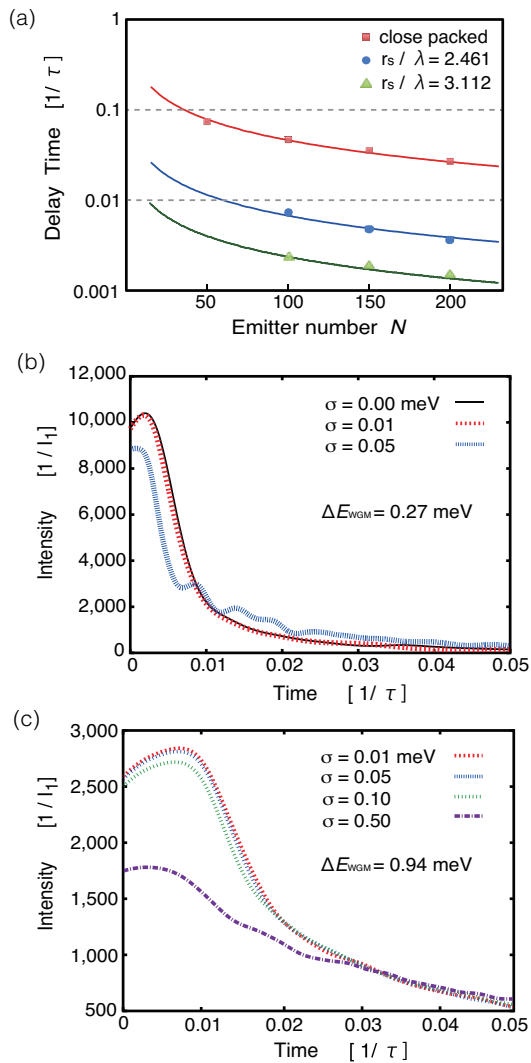}
\caption{(a) Delay times  of the synchronization $\tau_{d}$ in the case of close packed emitters and of those on the polystyrene spheres. The red line represents the analytical result $\tau^{\rm DS}_{d}/\tau=(\ln N)/N$, whereas the other two lines represent constant multiples of the red line to fit the plots. (b) Influence of the resonant-level distribution for the $N=100$ emitters. The radii of the polystyrene spheres are $r_s/\lambda =3.112$. (c) The same plot for $r_s/\lambda=2.461$. Here, $\sigma$ is the standard deviation from the peak resonant energies $3.0 {\rm eV}$, and $\Delta E_{\rm WGM}$ is the full width of the corresponding WGM.}
\label{syncro} 
\end{figure}

Nevertheless, the demonstrated fluorescence exhibits somewhat different behavior from Dicke superfluorescence scaled by the delay time  $\tau^{\rm DS}_{d}=\tau (\ln N)/N$~\cite{tSFr,tSF1}. The initial increase of the fluorescent intensity undoubtedly indicates the accelerated synchronization; the delay time follows the same scaling law as in Dicke superfluorescence, i.e., $\tau_{d}/\tau=\alpha (\ln N)/N$ [see Fig.~\ref{syncro}(a)]. The factors of proportionality are $\alpha=0.14$ for $\eta=0.06$ ($r_s/\lambda =2.461$), and $\alpha=0.05$ for $\eta=0.2$ ($r_s/\lambda =3.112$).  On the other hand, after achieving synchronization ($t>\tau_{d}$), the intensity shows the near-exponential decay, of which the time constant decreases with $\eta$. This is because, once the synchronization is achieved, the WGMs no longer mediate the emitters and the fluorescence decay is governed by the faster process: the photoemission via the WGM resonator. We can see from these results that nonuniform propagations of mediators play the key role in evaluating synchronization dynamics in open systems.

Finally, we mention a possible setup to verify the synchronized fluorescence. One of the candidates consists of dye molecules conjugated with proteins that coat a polystyrene microsphere~\cite{Gartia}. The substantive dyes must have some dispersion in the resonant energy. Then, we plot in Figs.~\ref{syncro}(a) and \ref{syncro}(b) the fluorescent dynamics when the $N=100$ emitters possess the distributed energies with standard deviation $\sigma$. Although the intensity drops considerably for $\sigma > \Delta E_{\rm WGM}/6$, the dispersion is not influential for smaller $\sigma$. Here, $\Delta E_{\rm WGM}$ is the full spectral width of the WGM. This can be understood by considering the characteristics of the standard deviation, whereby over $99.9 \%$ of the constituents are in the range between $\pm 3\sigma$ from the peak. Therefore, among the dye molecules coating the microsphere, only those whose resonant energies are covered by the WGM spectra participate in the synchronization. However, it seems feasible to synchronize hundreds of the qualifying emitters in reality, because $10^5$-$10^6$ molecules are tethered to the sphere~\cite{Gartia}.

To conclude, we have proposed a new scheme for examining synchronization of two-level emitters. As long as Green function for the surrounding environment is obtained analytically or numerically, the scheme enables us to simulate synchronizations in arbitrary systems, and helps to design a synchronizing system. As one example of design, we investigated emitters on a polystyrene microsphere, and demonstrated peculiar synchronization dynamics. The emitters can overcome the retardation effect and develop cooperative correlations even when their mean separation exceeds the radiation wavelength. Here, the WGMs combine two roles: mode resonator and synchronization mediator. Such synchronization dynamics can never be examined without incorporating the detailed information such as the emitters' locations and the mediators' propagations. As  ``designed synchronization'' can provide a direction to building original functions of many-body systems, our results offer a new paradigm in science and engineering.

We thank M. Enokida for useful discussions. This work was supported by JSPS KAKENHI Grant No. JP16H06504 in Scientific Research on Innovative Areas ``Nano-Material Optical-Manipulation,'' Grant-in-Aid for Challenging Exploratory Research No. 15K13505, and Grant-in-Aid for Scientific Research (C) No. 26400320.

\end{document}